\documentclass[12pt]{article}
\usepackage{times}
\usepackage{amsmath}
\usepackage{amssymb}
\usepackage[pdftex]{graphicx} 
\usepackage{epstopdf}         
\usepackage{ITC}
\usepackage{caption}
\usepackage{booktabs}
\usepackage{ifpdf}
\usepackage{subcaption}
\ifpdf
\usepackage[pdftex]{graphicx} 
\usepackage{epstopdf}         
\else
\usepackage{graphicx}         
\fi

\title{\Large \vspace{-0.5in}
\MakeUppercase{Elephant-Human Conflict Mitigation: An Autonomous UAV Approach}}
\author{Weiyun Jiang\thanks{These two authors contributed equally.},  Yukai Yang\footnotemark[1]  \\
    \normalsize Electrical and Computer Engineering\\
    \normalsize University of California, Santa Barbara, CA\\
    \normalsize \{weiyunjiang, yyang01\}@ucsb.edu\\[6pt]
    Faculty Advisor: \\
    Yogananda Isukapalli}

\date{}

\begin{document}

\maketitle 

\section{\MakeUppercase{Abstract}}

\noindent
Elephant-human conflict (EHC) is one of the major problems in most African and Asian countries. As humans overutilize natural resources for their development, elephants' living area continues to decrease; this leads elephants to invade the human living area and raid crops more frequently, costing millions of dollars annually. To mitigate EHC, in this paper, we propose an original solution that comprises of three parts: a compact custom low-power GPS tag that is installed on the elephants, a receiver stationed in the human living area that detects the elephants' presence near a farm, and an autonomous unmanned aerial vehicle (UAV) system that tracks and herds the elephants away from the farms. By utilizing proportional–integral–derivative controller and machine learning algorithms, we obtain accurate tracking trajectories at a real-time processing speed of 32 FPS. Our proposed autonomous system can save over 68\% cost compared with human-controlled UAVs in mitigating EHC.

\section{\MakeUppercase{Introduction}}

\noindent
Elephant-human conflict (EHC) has been one of the most significant problem in most African and Asian countries. EHC is extremely prevalent because nearly 1.2 billion people in the world live in African and Asian elephant range countries~\cite{shaffer2019human}. Crop raiding is one of the most common types of EHC~\cite{sitati2003predicting}. C. Mackenzie, et al. found that elephants around Kibale National Park, Uganda, damage various crops, such as maizes, beans, sweet potatoes, and so on, and lead to over US\$3500 of total economic loss in a village of 145 households during 6 months~\cite{mackenzie2012elephants}. These crops are the main source of income for these villagers in the village, where the median household capital asset wealth was only US\$5033. Crop raiding causes huge economical loss to not only small vegetation, but also large plantation. Riau, the largest palm oil producing province in Indonesia, loses millions of dollars during to crop-raiding~\cite{perera2009human}. Besides the tremendous economic loss, EHC also leads to casualties of both human and elephants. EHC in India leads to approximately 400 human deaths and 100 elephant deaths annually~\cite{gulati2021human}.

Traditional methods of mitigating EHC are limited to human guarding, fire, beating drums, scare shooting, dogs, and so on~\cite{sitati2006assessing}. These methods are not cost-effective and efficient because almost all these traditional methods require expensive human labor. In addition, elephants begin to get used to these methods after people have used them many times. Recently, researchers  in~\cite{raya2017small} founds that UAVs can mimic honeybees' humming sound, which is known to annoy elephants. These small bees love to sting elephants' sensitive areas such as eyes, ears, and noses~\cite{raya2017small}. N. Hahn, et al. have also proven the effectiveness of drones by hiring rangers to manually control UAVs to herd elephants away from the farms~\cite{hahn2017unmanned}, in this paper we build on this idea and propose an autonomous UAV system.
\begin{figure}[t]
\begin{subfigure}{0.29\textwidth}
  \centering
  \vspace{-4mm}
  \captionsetup{justification=centering}
  \includegraphics[width=0.9\linewidth]{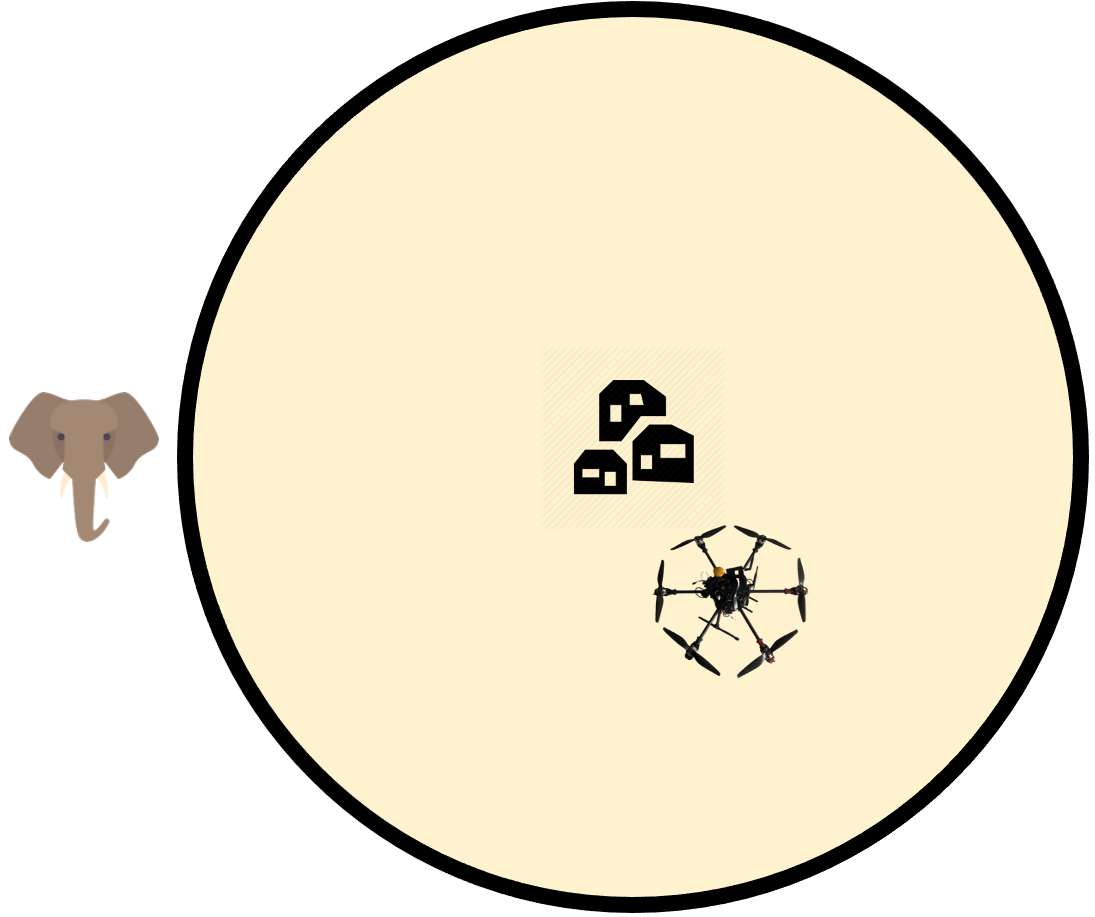}
  \caption{Elephant at the border}
\end{subfigure}\hfill
\begin{subfigure}{0.25\textwidth}
  \centering
  \captionsetup{justification=centering}
  \includegraphics[width=0.9\linewidth]{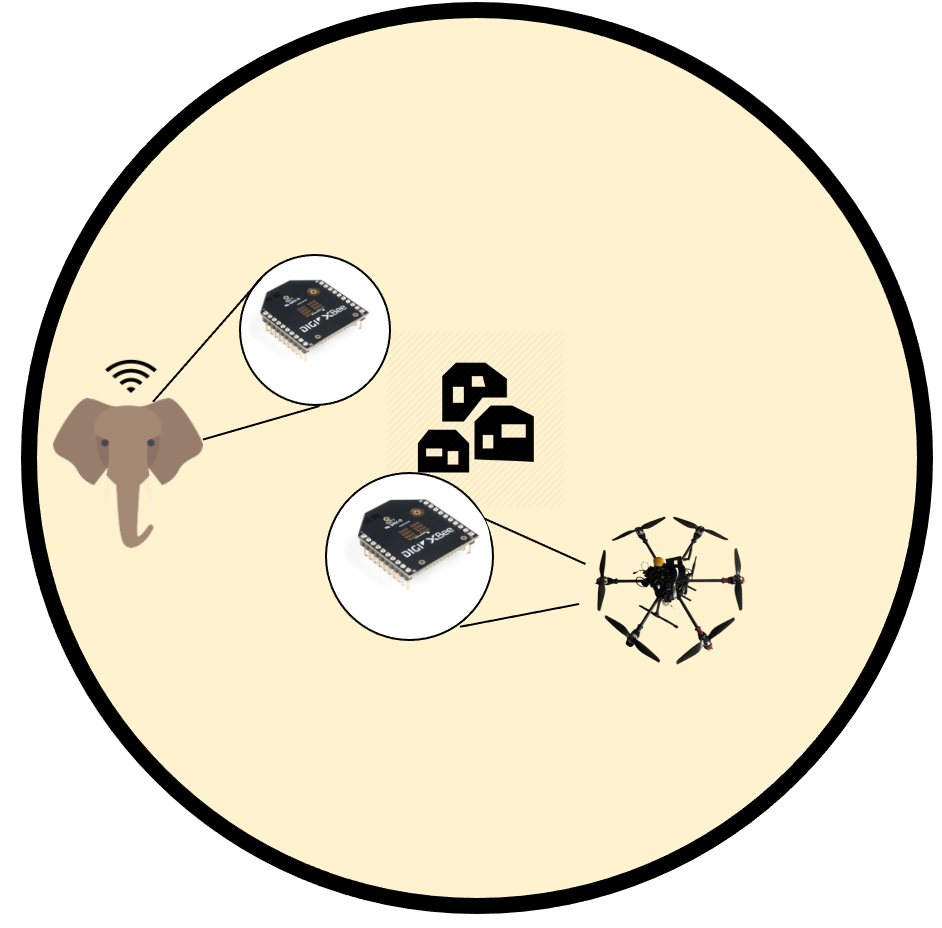}
  \caption{Elephant invades the farm}
\end{subfigure}\hfill
\begin{subfigure}{0.25\textwidth}
  \centering
  \captionsetup{justification=centering}
  \includegraphics[width=0.9\linewidth]{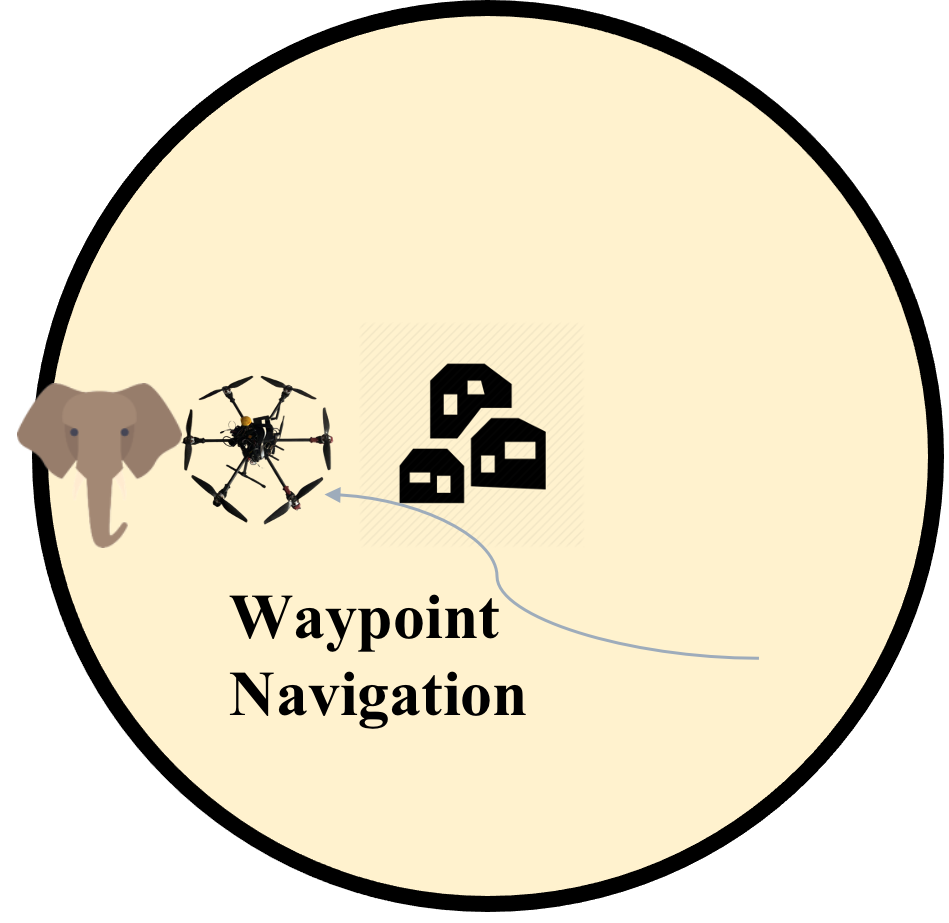}
  \caption{UAV navigates to the elephant}
\end{subfigure}
\begin{subfigure}{0.25\textwidth}
  \centering
  \captionsetup{justification=centering}
  \includegraphics[width=0.9\linewidth]{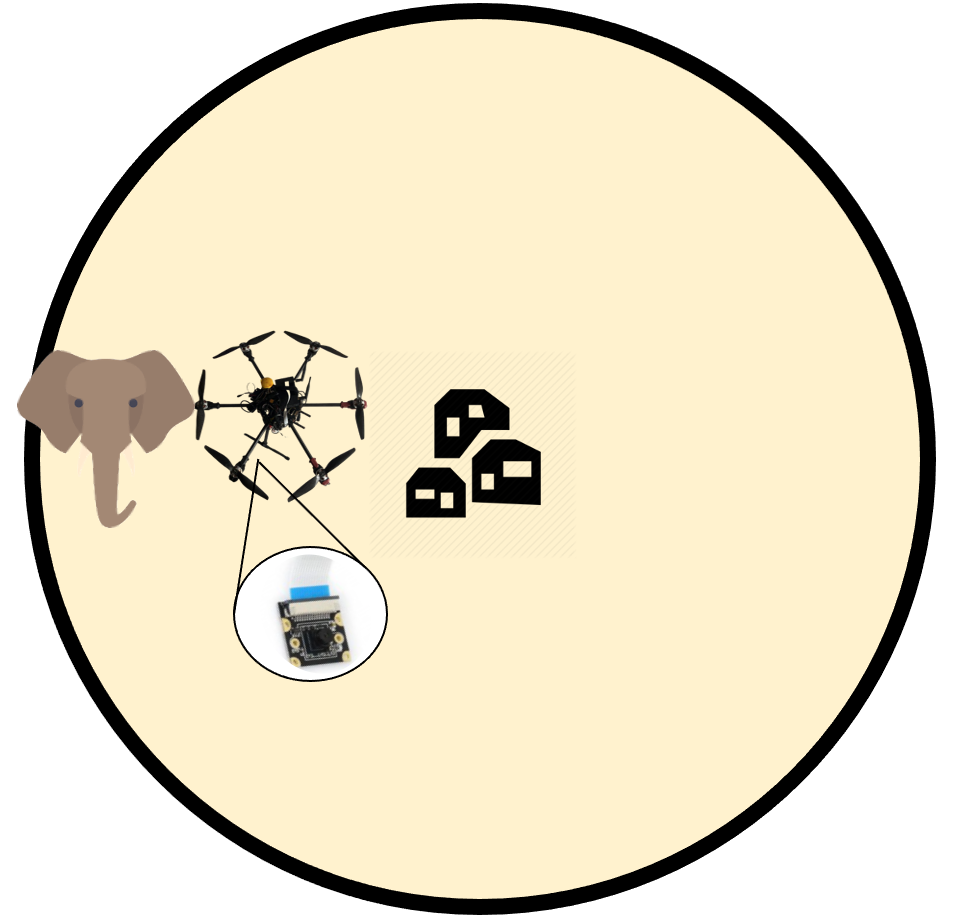}
  \caption{Camera turns on}
\end{subfigure}\hfill
\begin{subfigure}{0.3\textwidth}
  \centering
  \captionsetup{justification=centering}
  \includegraphics[width=0.9\linewidth]{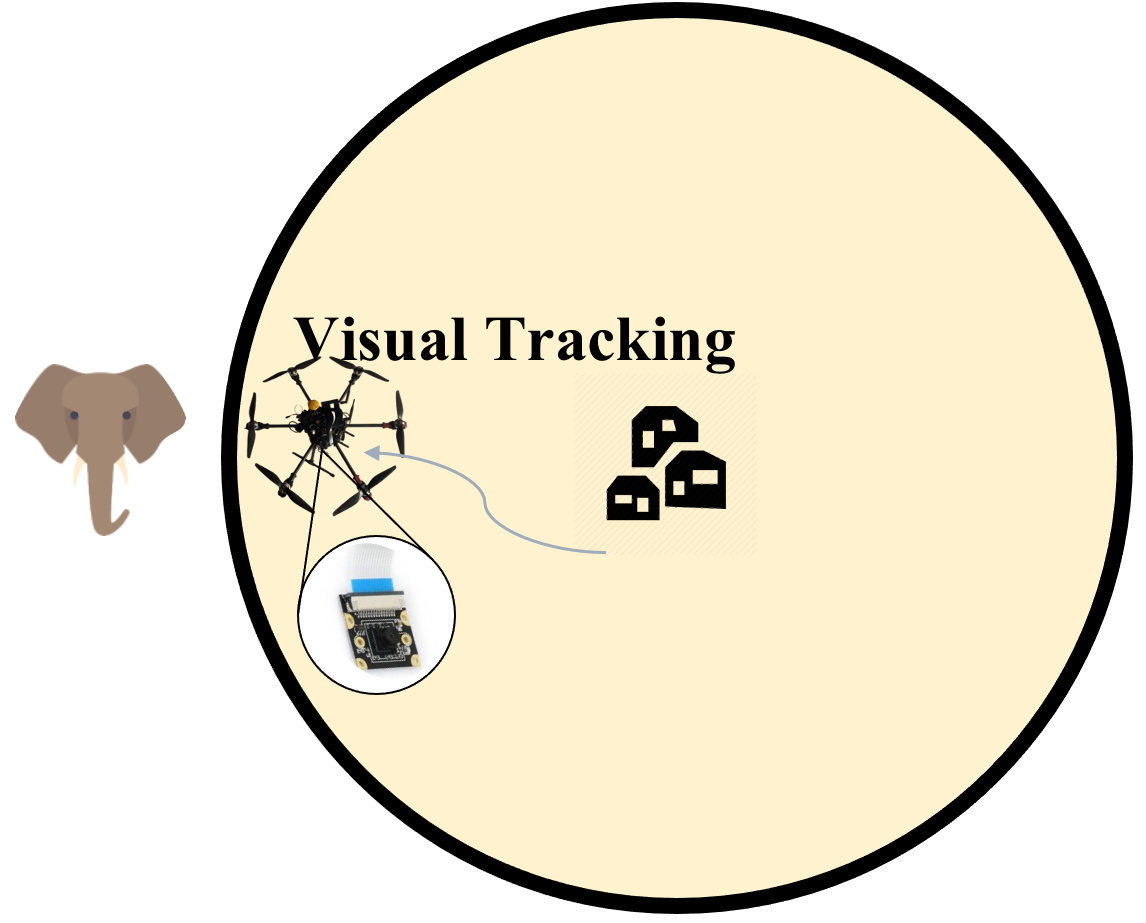}
  \caption{Elephant herded away}
\end{subfigure}
\caption{Our Autonomous UAV Illustration}
\label{fig:system_illus}
\end{figure}
\subsection{Paper Contributions}
Our contribution in this paper can be summarized as follows:
\begin{itemize}
    \item Our solution comprises of three parts: a compact custom low-power GPS tag installed on the elephants, a receiver stationed in the human living area that detects the elephants' presence near a farm or village, and an autonomous UAV system that tracks and herds the elephants away from the farms. 
    \item By utilizing PID controller and machine learning techniques, we obtain accurate tracking trajectories at a real-time processing speed of 32 FPS.
    \item Our proposed autonomous system can save over 68\% cost compared with human-controlled UAVs in mitigating EHC.
\end{itemize}

\section{Overall Approach}
\label{sec:Approaches}
Our autonomous UAV solution can be summarized as follows:
\begin{itemize}
    \item To begin with, we place a GPS tag on the elephant (ankle or neck) by tranquilizing it.  And then, a ground base is set up on the farm, where protection is needed from elephants' invasions. The ground base will emit signals that the GPS tag will pick up when the signal is strong enough. So the range of the signal from the ground base defines the perimeter of the protected area we want to prevent elephants from entering. 
    \item When the elephant with the GPS tag goes into the range of the signal of the ground base, the GPS tag will be woken up from the sleep mode, and it will start to send out real-time coordinates of the elephant. 
    \item Once the UAV (situated in the farm) receives the coordinates of the elephants, it navigates to the location of the elephant. 
    \item When the UAV is closer to the coordinates of the elephant, the onboard camera of the UAV is enabled, and the live video feed is then processed by vision algorithms on NVIDIA Jetson Nano to track the elephant effectively. 
    \item Finally, the drone will herd the elephant away from the ground base and prevent elephants from damaging crops since the UAV emits sounds similar to honeybees that elephants are afraid of. Elephants tend to stay away from bees because these small bees love to sting elephants' sensitive areas, such as eyes and ears~\cite{vollrath2002african}.
\end{itemize}


Figure~\ref{fig:system_illus} illustrates the process of our proposed autonomous UAV system in more details. The yellow shaded circle represents the region of the signal covered by the ground station. And the three little houses in the middle of the circle represent the ground station. In addition, the yellow shaded circle is the region where the signal from the base station will trigger and wake up the GPS tag on the elephant. Initially, when the elephant is outside the yellow region, which is the protected region of the farm, the drone is on the ground and ready to take off. 

When the elephant comes into the protected region around the ground base, the Xbee module receives the signal from the ground base, which triggers and wakes up the GPS tag on the elephant. Then the GPS coordinate of the elephant is sent out by the Xbee module on the GPS tag and is then received by the Xbee module on the drone. The drone immediately takes off and uses waypoint navigation to fly to the coordinate of the elephant.

When the drone is close enough to the position of the elephant, the onboard camera will turn on. GPS signals only provide a coarse location of the elephant; the vision algorithm enables the drone to track the elephant’s movement in a more fine-grained fashion. Finally, the drone maneuvers and herds the elephants until they are outside the protected area.

\section{Hardware Design}
\begin{figure*}[h]
  \begin{center}
    \includegraphics[width=0.58\textwidth]{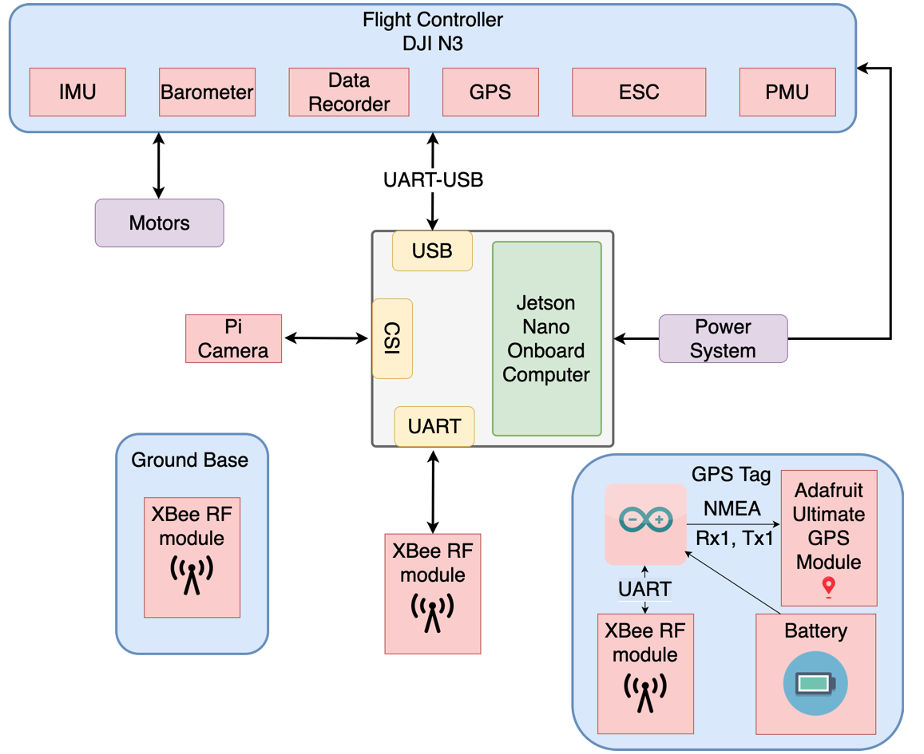}
  \end{center}
  \vspace{-5mm}
  \caption{System Block Diagram}
  \label{fig:system_block_diagram}
\end{figure*}

In this section, we discuss our hardware building blocks shown in Figure~\ref{fig:system_block_diagram}, which include the custom UAV and the GPS tag installed on elephants.

Our UAV is a standard hexacopter with six motors and electronic speed controllers (ESC) complemented by an onboard computer, a camera, and a communication module.

\begin{figure}[h]
\centering
\begin{subfigure}{.45\textwidth}
  \vspace{-10mm}
  \begin{center}
    \includegraphics[width=0.7\textwidth]{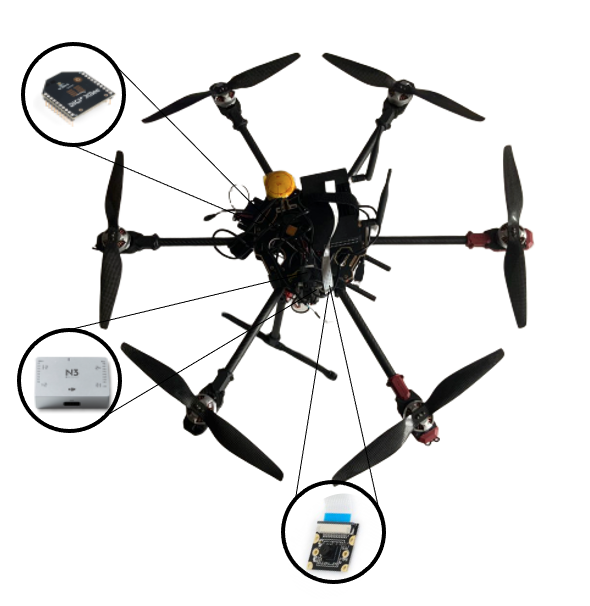}
  \end{center}
  \vspace{-4mm}
  \caption{UAV Prototype}
  \label{fig:uav}
\end{subfigure}%
\begin{subfigure}{.45\textwidth}
  \begin{center}
    \includegraphics[width=\textwidth]{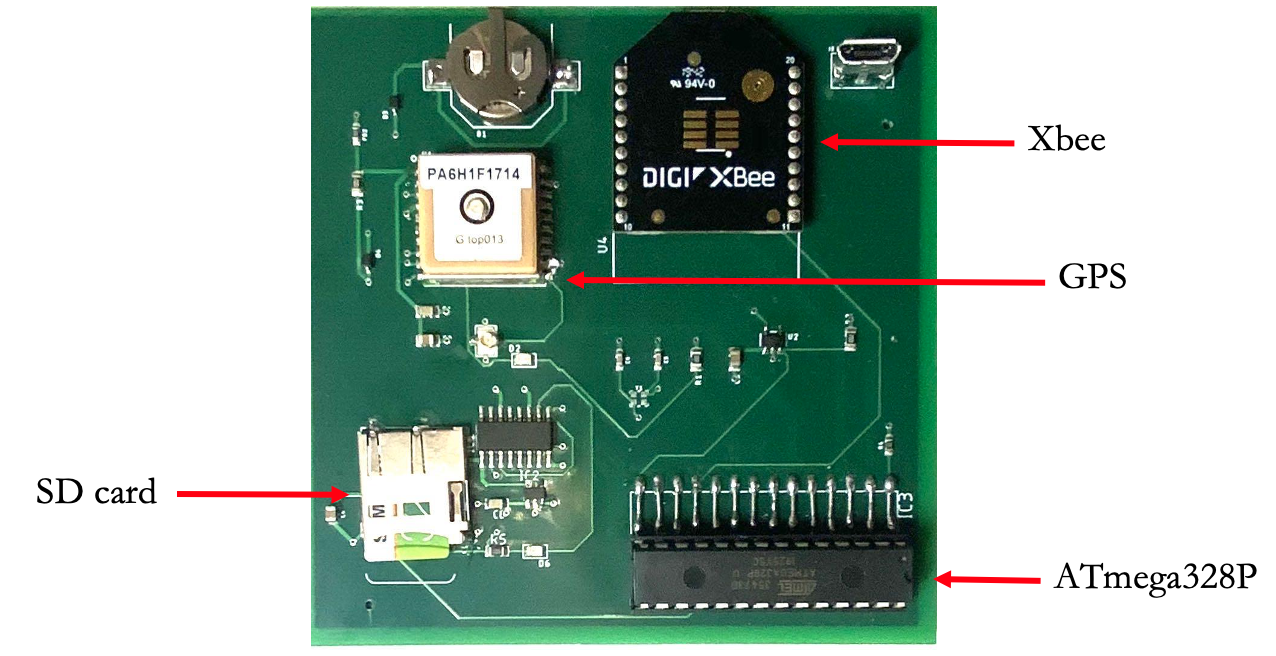}
  \end{center}
  \caption{GPS Tag PCB}
  \label{fig:pcb}
\end{subfigure}
\vspace{-2mm}
\caption{Hardware Design}
\label{fig:hardware}
\end{figure}

We select Jetson Nano as the onboard processor due to its low power consumption and compactness. Further, its integrated NVIDIA GPU allows us to perform neural network inference with hardware acceleration. Figure~\ref{fig:uav} shows our completed UAV prototype, with the DJI N3 flight controller, Xbee RF module, and the camera mounted. DJI N3 has multiple built-in sensors such as GPS and IMU that support stable waypoint navigation and control. The Xbee RF module is responsible for receiving the location of the elephants sent by the GPS tag. To ensure our UAV can complete the round-trip herding task, we utilize a high capacity 12000 mAh LiPo battery to power the entire UAV system. The LiPo battery allows flight time as long as 20 minutes and enables the UAV to travel up to 4 miles without recharging.

As shown in Figure~\ref{fig:pcb}, our GPS tags are made of four major components: a Xbee RF module, a GPS module, a SD card reader and a ATmega328P micro-controller. Further, we also include a button cell for the GPS module. This button cell prevents the GPS module from shutting down completely. It takes more than 10 minutes for the GPS module to find a fixed location after shutting down completely. Thus, the included button cell helps the GPS module to find a fixed and accurate location coordinate quickly.

The GPS tag is a 4-layer board of 83 mm $\times$ 83 mm and weighs only 26.3 g. Thus, elephants will barely notice the extra weight if we install our GPS tags on them. The electric power consumption of our GPS tag is 10.5 mW. If we power our GPS tag with a 10000 mAh battery bank, our GPS tags can last for about 200 days on the elephants.

\section{Algorithm Design}
In this section, we discuss our algorithms that run onboard the UAV and how we meet the real-time processing constraint. Our onboard processing has two major subcomponents, visual tracking and control. When we launch our control program, it will spawn the tracker as a separate process and communicate with it through a socket. The tracker will send the bounding box coordinates of the elephant to the control program at each frame. Upon receiving the bounding box coordinates, the control program will generate control signals for the UAV.


\subsection{Detection} 
The rapid advancement of Convolutional Neural Network (CNN) has brought object detection to a new level. SORT \cite{Bewley2017}, a similar framework to ours, adopts Faster-RCNN, an accurate yet expensive two-stage model. To deal with the compute constraint, we utilize SSD \cite{Liu2015}, a single-stage and highly efficient detector. We select $300\times300$ pixels as the input size and Inception V2 as the backbone. The advantage of SSD is that when quantized using TensorRT, we obtained near real-time inference speed (around 25 FPS) in our environment. Since we are interested in persons, vehicles, and wildlife, we trained the model on COCO \cite{Lin}, a widely used detection dataset.

Our experiments found that even though SSD has comparable accuracy with two-stage detectors on large and medium-sized objects, it struggles to detect objects with smaller pixel-size consistently. The reason is that SSD only regresses bounding boxes on downsampled feature maps in the last few layers of the CNN backbone. However, it is impractical for us to increase the input size or tweak the SSD architecture due to the hardware constraint. 

Therefore, we utilize tiling as a workaround. We divide the $1280\times720$ video frame into 6 $300\times300$ tiles. The tiles overlap with each other by $25\%$ to cover objects near the edges. Ideally, we should be able to process all 6 tiles in parallel through batching. However, we are once again limited by the compute constraint of our environment. As a result, we have to process the tiles sequentially, and we introduce a simple attention mechanism by weighting the tiles containing a large number of objects more than empty tiles. To refrain from starvation, we also adopt an aging mechanism that keeps track of the number of frames since the tile was last processed. Lastly, we set the confidence threshold of SSD to $0.5$ to filter out possible false positives.

\subsection{Tracking}
\begin{figure}[t]
\centering
\begin{subfigure}{.5\textwidth}
  \centering
  \vspace{15mm}
  \includegraphics[width=0.8\textwidth]{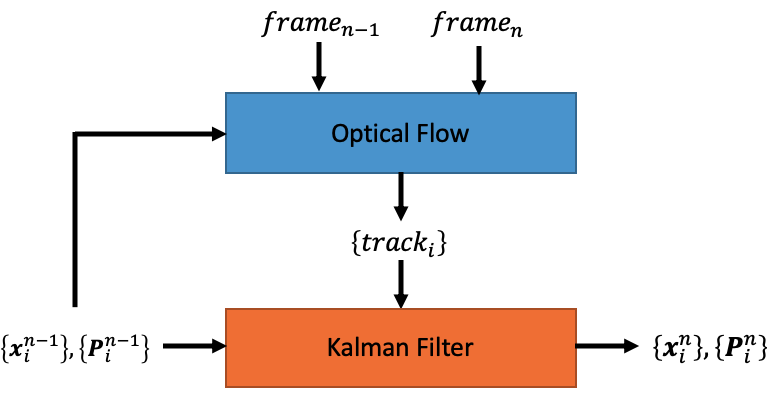}
  \caption{Our tracking module}
  \label{fig:tracking_module}
\end{subfigure}%
\begin{subfigure}{.5\textwidth}
  \centering
  \includegraphics[width=0.8\textwidth]{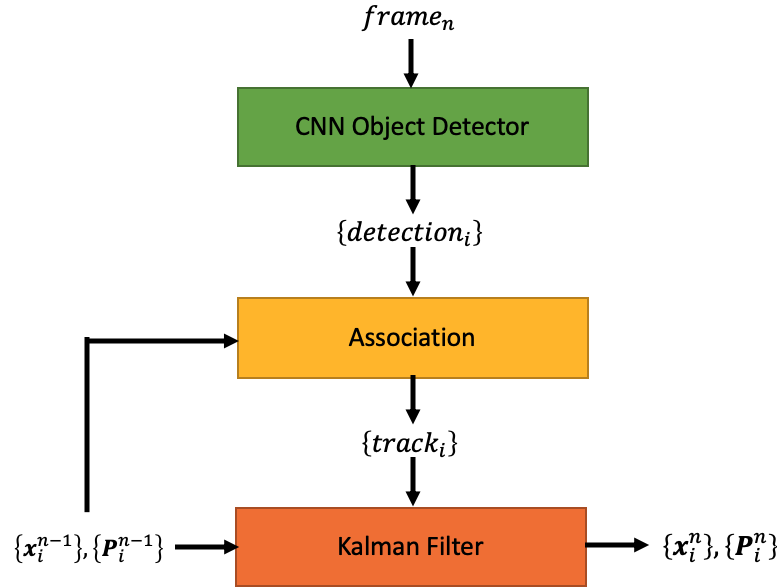}
  \caption{Our detection module}
  \label{fig:detection_module}
\end{subfigure}
\vspace{-2mm}
\caption{Tracker Design}
\label{fig:algorithm}
\end{figure}
SORT~\cite{Bewley2017} assumes that the detector runs at every frame and only relies on the Kalman filter to process the detections, which is not realistic on an edge device. Furthermore, we can process a tile $1 / 6$ of the time on average, or worse, when the detector cannot achieve real-time inference because of sequential tile processing. Therefore, we need to track objects from frame to frame when SSD does not process a specific tile for a large number of frames. 

Most correlation and CNN-based trackers are thus unfit for this purpose since their speed scales poorly with the number of objects. Instead, we utilize optical flow, a lightweight classical algorithm that tracks feature points on the objects. For each object, we use ShiTomashi corners and sample a fixed density of them inside the bounding box. We model bounding box transformation as affine and compute a transformation matrix from optical flow feature matches to estimate a new bounding box for the next frame. We also estimate a homography transformation from background feature matches, which is helpful to compensate for camera motion during Kalman filtering \cite{White}. 

\subsection{Motion Model}
\label{sec:motion_model}
We adopt the standard Kalman filter and a constant velocity model to alleviate optical flow's undesirable performance when the targets are partially occluded. Since Kalman filter models the motion of objects, the next bounding box locations can be roughly predicted based on velocity when track measurements are incorrect (occluded).
Our Kalman filter state for each object is defined on the eight dimensional state space:
\begin{equation}
\boldsymbol{x}=\begin{bmatrix}x_{tl} & y_{tl} & x_{br} & y_{br} & \dot{x}_{tl} & \dot{y}_{tl} & \dot{x}_{br} & \dot{y}_{br} \end{bmatrix}^{T}
\end{equation}
where $x_{tl}$, $y_{tl}$ represent horizontal and vertical pixel position of the top left corner of the bounding box, $x_{br}$, $y_{br}$ represent horizontal and vertical pixel position of the bottom right corner of the bounding box. $\left(\dot{x}_{tl}, \dot{y}_{tl}, \dot{x}_{br}, \dot{y}_{br}\right)$ are the corresponding velocities. The track bounding boxes are used as measurements to update the state of Kalman filter. Note that our measurement space only consists of pixel positions, $\left(x_{tl}, y_{tl}, x_{br}, y_{br}\right)$.

Using only the pixel positions of bounding box corners instead of its width and height like \cite{Bewley2017} and \cite{Wojke} enables us to easily transform Kalman filter state to compensate for camera motion. However, the two corners of the bounding box are modeled independently, which can cause the bounding box to drift away from its groundtruth size over time. To tackle this issue, we update the positions of the two corners jointly when applying the constant-velocity prediction:
\begin{equation}
    x_{tl} = x_{tl} + (\lambda\dot{x}_{tl} + (1 - \lambda)\dot{x}_{br}) \Delta t
\end{equation}
\begin{equation}
    y_{tl} = y_{tl} + (\lambda\dot{y}_{tl} + (1 - \lambda)\dot{y}_{br}) \Delta t
\end{equation}
where $\lambda$ is the coupling factor that measures the correlation between the two corners. We update $x_{br}$ and $y_{br}$ in a similar fashion. Through empirical experiment, we set $\lambda = 0.6$. 
\begin{figure*}[t]
  \begin{center}
    \includegraphics[width=\textwidth]{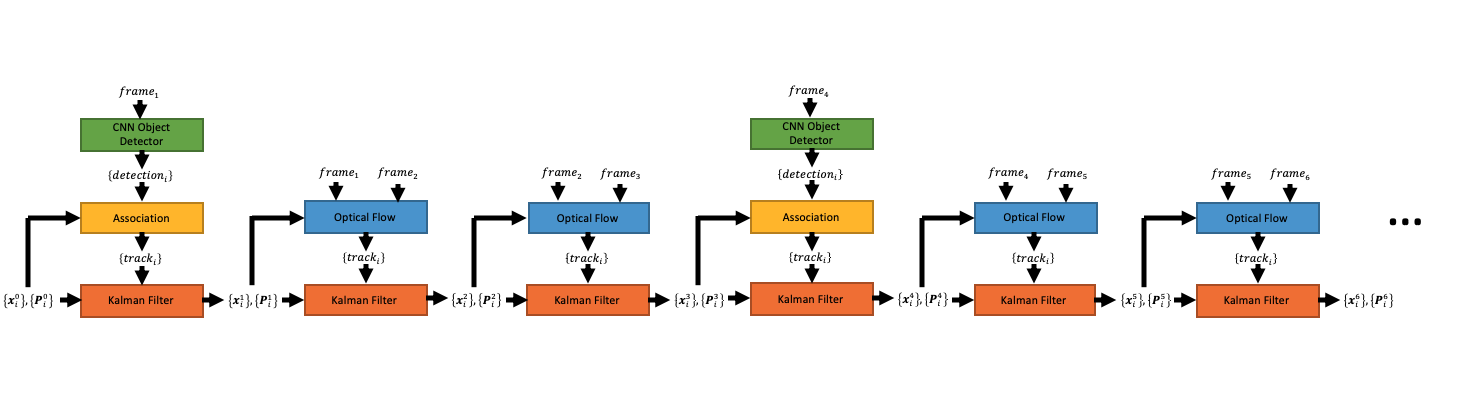}
  \end{center}
  \vspace{-10mm}
  \caption{Unrolled Tracker Pipeline ($N = 3$)}
  \label{fig:pipeline}
\end{figure*}
\begin{figure}[t]
  \begin{center}
    \includegraphics[width=0.4\textwidth]{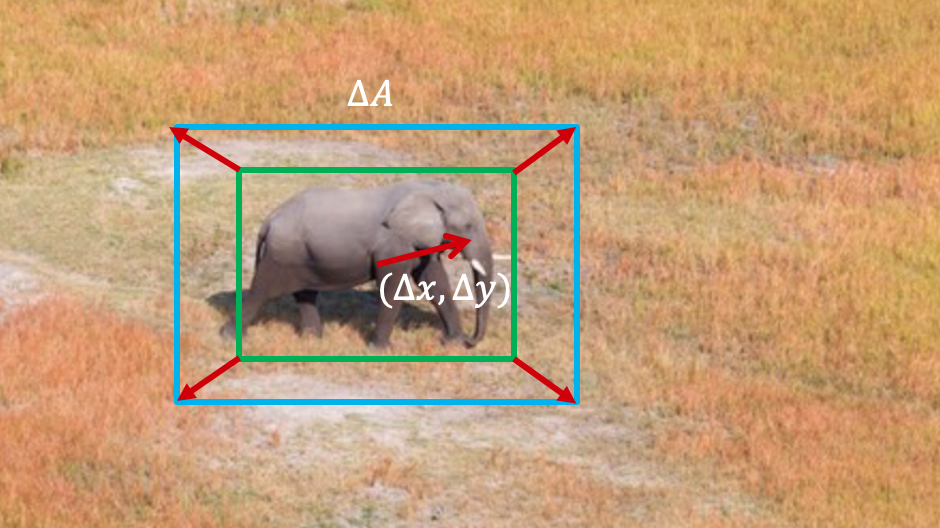}
  \end{center}
    \vspace{-5mm}
  \caption{PID Control Errors}
  \label{fig:pid}
\end{figure}
\subsection{Data Association}
\label{sec:data_association}
To associate detections to tracks, we make use of both the state $\boldsymbol{x}$ and the estimated covariance of the state $\boldsymbol{P}$. We partially follow the approach in Deep SORT \cite{Wojke} by computing the squared Mahalanobis distance between the Kalman filter state and the detection:
\begin{equation}
d(i, j)=\left(\boldsymbol{d}_{j}-\boldsymbol{\hat{x}}_{i}\right)^{\mathrm{T}} \boldsymbol{\hat{P}}_{i}^{-1}\left(\boldsymbol{d}_{j}-\boldsymbol{\hat{x}}_{i}\right)
\end{equation}
where $(\boldsymbol{\hat{x}}_{i}, \boldsymbol{\hat{P}}_{i})$ is the i-th track state $(\boldsymbol{x}_{i}, \boldsymbol{P}_{i})$ projected onto the measurement space and $\boldsymbol{d}_{j}$ is the j-th bounding box detection. The Mahalanobis distance factors in state uncertainty by measuring how many
standard deviations the detection is away from the mean track position.

We filter out unlikely associations with large $d(i, j)$ and threshold Intersection Over Union (IOU) at 0.3. IOU measures the percentage of overlap between the detection and track bounding boxes. The Hungarian algorithm is used obtain the optimal $d(i, j)$ assignment. Finally, the remaining unassociated detections are registered as new tracks.
\begin{table}[t]
\centering
\caption{Cost Summary}
\begin{tabular}{@{}ccc@{}}
\toprule
\multicolumn{1}{l}{Item}                       & \begin{tabular}[c]{@{}c@{}}Unit Cost (US\$)\end{tabular} & \begin{tabular}[c]{@{}c@{}}Annual Cost (US\$)\end{tabular} \\ \midrule
\multicolumn{1}{l}{Custom UAV Kits}            &                                                            &                                                              \\
Camera                                         & 23.5 ($\times3$)                                   
& 70.5                                                         \\
Flight Controller                       & 419 ($\times3$)                                        
& 1257                                                            \\
Jetson Nano                                    & 100 ($\times3$)                                      
& 300                                         \\
Frame                                         & 170 ($\times3$)                                       
& 510                                       \\
Motors                                         & 180 ($\times3$)                                       
& 540                                        \\
ESCs                                            & 126 ($\times3$)                                       
& 378                                        \\
Propellers                                     & 60 ($\times3$)                                       
& 180                                         \\
LiPo Battery                                            & 83 ($\times3$)                                
& 249                                        \\
\multicolumn{1}{l}{\textit{Subtotal Kits}}     & 1161.5                                       
& 3484.5                                         \\
\multicolumn{1}{l}{Custom GPS Tags}            &                                                            &                                                              \\
GPS Module                                     & 13 ($\times10$)                                                   & 130                                                          \\
Xbee Module                                    & 25 ($\times10$)                                                   & 250                                                          \\
SD Card                                        & 14 ($\times10$)                                                   & 140                                                          \\
Atmega328p                                     & 2.5 ($\times10$)                                                  & 25                                                           \\
Battery Bank                                   & 15 ($\times10$)                                                   & 150                                                          \\
\multicolumn{1}{l}{\textit{Subtotal GPS Tags}} & 69.5                                                       & 695                                                          \\
\multicolumn{1}{l}{Elephant Anesthesia}        & \multicolumn{1}{l}{}                                       & 500                                                          \\
\textbf{Total}                                        & \multicolumn{1}{l}{}                                       & \textbf{4679.5}                                       \\ \bottomrule
\end{tabular}
\label{tab:cost}
\end{table}
\subsection{Technical Approach} 
In this section, we provide a detailed overview of our visual tracking algorithm. Figure~\ref{fig:tracking_module} shows our tracking module, which is run at almost every frame. For each target, optical flows takes as input the previous frame, the current frame, and Kalman filter's last state $\left(\boldsymbol{x}_{i}^{n-1}, \boldsymbol{P}_{i}^{n-1}\right)$ to estimate a bounding box for the target in the current frame and outputs it as a new track. Each track is associated with a track ID to identify the track. After Kalman filter predicts a new state based on the constant-velocity model, the new track will be used to update (correct) Kalman filter's new state to produce a final state output, $\left(\boldsymbol{x}_{i}^{n}, \boldsymbol{P}_{i}^{n}\right)$.

Figure~\ref{fig:detection_module} shows our detection module, which is run every $N$ frames due to its large latency. The SSD detector takes as input the current frame and outputs a set of detections. Each detection will be associated to a track using the approach in \ref{sec:data_association} Similarly, each track output by association will be used to update Kalman filter's predicted state to produce a final state output, $\left(\boldsymbol{x}_{i}^{n}, \boldsymbol{P}_{i}^{n}\right)$.

Figure~\ref{fig:pipeline} shows our unrolled tracker pipeline that combines both modules, where the detection module is inserted every $3$ frames starting from $frame_{1}$.

\subsection{Control}

We incorporate  proportional–integral–derivative (PID) controllers for yaw, pitch, and roll velocities based on 2D bounding boxes output by our tracker. In Figure \ref{fig:pid}, we define three control errors, $\Delta x$, $\Delta y$, and $\Delta A$. $\Delta x$ and $\Delta y$ are the horizontal and vertical distances, respectively, between the centroid of the bounding box and that of the entire video frame. We also define a reference bounding box (blue) with a fixed area. $\Delta A$ is the area difference between the current bounding box and the reference bounding box, which roughly indicates a change in spatial distance. We use $\Delta x$ to control yaw and roll, while $\Delta y$ and $\Delta A$ are used to perform PID on the pitch of the UAV.
\vspace{-1mm}
\section{\MakeUppercase{Results}}
\label{sec:Results}

\begin{figure}[t]
  \begin{center}
    \includegraphics[width=0.65\textwidth]{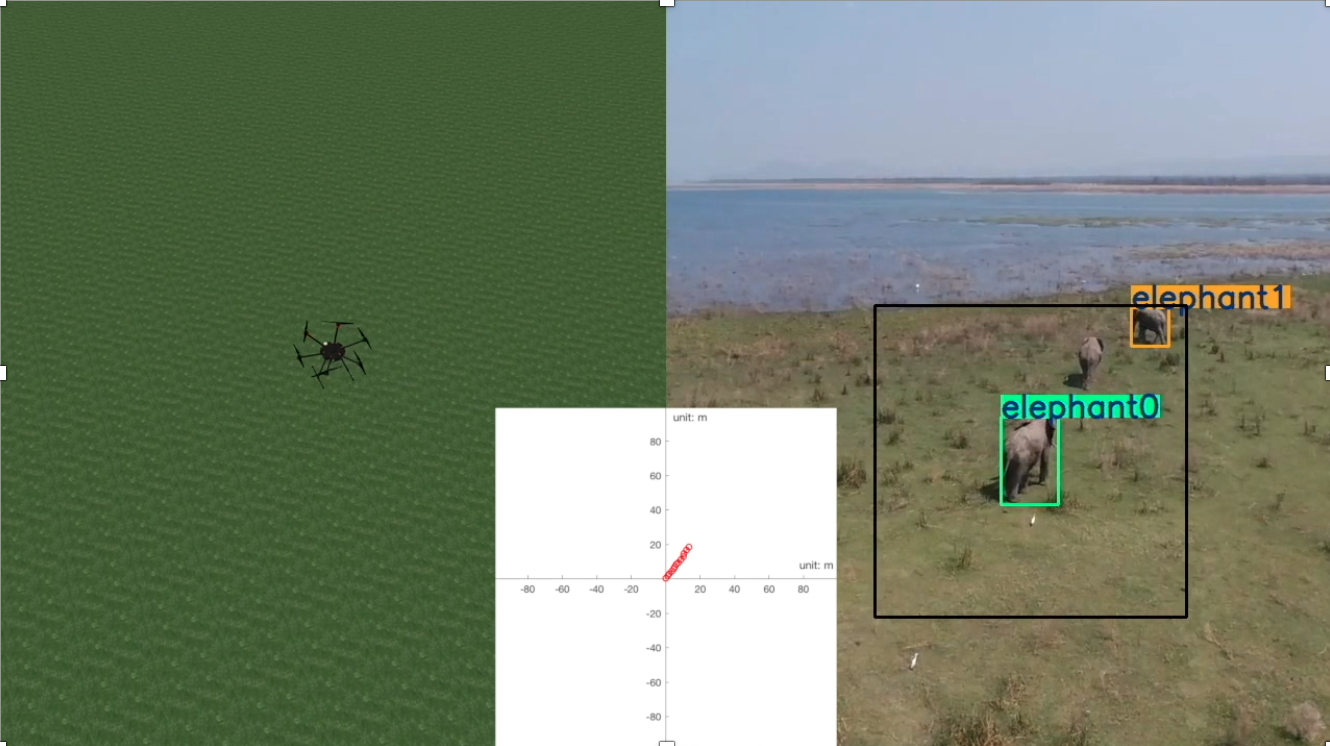}
  \end{center}
  \caption{Simulation Result}
  \label{fig:sim}
\end{figure}

The proposed tracker achieves high tracking precision at 32 FPS on Jetson Nano, which is fast enough to control the UAV in real-time. To evaluate the control algorithm, in Figure~\ref{fig:sim}, we simulate the flight trajectories in the DJI Flight Simulator. To retrieve the ground truth 3D trajectory, we project the 2D image coordinates of the elephant of interest (green bounding box) in the video to the ground plane. As a result, the trajectory of the UAV has 80\% match with the ground truth trajectory given an error margin of a 1-meter radius.

N. Hahn et al. investigated the cost of human-controlled UVAs on the borders of Tanzanian Parks, which is about US\$15,000 annually~\cite{hahn2017unmanned}. To cover the area of interest with our solution, we estimate 10 GPS tags and 3 UAVs are required on average per year. According to Table~\ref{tab:cost}, our proposed autonomous system only costs US\$4679.5 annually, which saves over 68\% cost compared with human-controlled UAVs in mitigating EHC.


\vspace{-1mm}
\section{\MakeUppercase{Conclusions}}
\label{sec:Conclusions}

\noindent
In this paper, we proposed an autonomous system that uses telemetry to track and herd elephants away from the villages. We design and implement all three parts of our solution: a compact custom low-power GPS tag, a receiver stationed in the human living area, and an autonomous UAV. Our real-time tracking algorithm achieves 32 FPS on the mobile edge device, NVIDIA Jetson Nano. And our proposed autonomous system is highly cost-effective compared with human-controlled solutions, saving more than 68\% per year. 

\bibliographystyle{ieeetr}
\bibliography{references}
\end{document}